\documentclass[useAMS,usenatbib,fleqn]{mn2e}

\usepackage{graphicx}
\usepackage{float}
\usepackage{epstopdf}
\usepackage{amsmath}
\usepackage{amssymb}

\usepackage{epsfig}
\usepackage{times}
\usepackage{color}

\def\gtsim {\gtrsim} 
\def\ltsim {\lesssim} 

\defcitealias{pt14}{TK14}
\defcitealias{pt15a}{TK15}

\title[Quantifying AGN-Driven Outflows]{ Quantifying AGN-Driven Metal-Enhanced Outflows in Chemodynamical Simulations}
\author[P.~Taylor and C.~Kobayashi]{Philip~Taylor$^1$\thanks{E-mail:p.taylor7@herts.ac.uk} and Chiaki~Kobayashi$^{1,2}$\\\\
$^1$Centre for Astrophysics Research, Science and Technology Research Institute, University of Hertfordshire, AL10 9AB, UK\\
$^2$Distinguished Visitor at Research School of Astronomy and Astrophysics, The Australian National University, Australia}

\begin{document}

\date{Accepted Received ; in original form}

\pagerange{\pageref{firstpage}--\pageref{lastpage}} \pubyear{}

\maketitle

\label{firstpage}


\begin{abstract}
We show the effects of AGN-driven outflows on the ejection of heavy elements using our cosmological simulations, where super-massive black holes originate from the first stars.
In the most massive galaxy, we have identified two strong outflows unambiguously driven by AGN feedback.
These outflows have a speed greater than $\sim 8000$ km\,s$^{-1}$ near the AGN, and travel out to a half Mpc with $\sim 3000$ km\,s$^{-1}$.
These outflows remove the remaining gas ($\sim 3$ per cent of baryons) and significant amounts of metals ($\sim 2$ per cent of total produced metals) from the host galaxy, chemically enriching the circumgalactic medium (CGM) and the intergalactic medium (IGM).
17.6 per cent of metals from this galaxy, and 18.4 per cent of total produced metals in the simulation, end up in the CGM and IGM, respectively.
The metallicities of the CGM and IGM are higher with AGN feedback, while the mass--metallicity relation of galaxies is not affected very much.
We also find `selective' mass-loss where iron is more effectively ejected than oxygen because of the time-delay of Type Ia Supernovae.
AGN-driven outflows play an essential role not only in quenching of star formation in massive galaxies to match with observed down-sizing phenomena, but also in a large-scale chemical enrichment in the Universe.
Observational constraints of metallicities and elemental abundance ratios in outflows are important to test the modelling of AGN feedback in galaxy formation.
\end{abstract}



\section{Introduction}
\label{sec:intro}

The importance of feedback from active galactic nuclei (AGN) has been underscored by the discovery of the relationship between the mass of the central black hole (BH) and the mass of the host galaxy bulge \citep{magorrian98,kormendyho13}, suggesting co-evolution of BHs and their host galaxies.
This has already been indicated from the similar shapes between the observed cosmic star formation history \citep[e.g.,][]{madau96} and the quasar space density \citep[e.g.,][]{schmidt95}. 
AGN feedback has been implemented in cosmological simulations, which provided an excellent agreement with the Magorrian relation { \citep[e.g.,][]{dimatteo08,sijacki14} }and a better reproduction of the cosmic star formation rates { (e.g., \citet[hereafter \citetalias{pt14}]{pt14}, \citealt{vogelsberger14}).}
The [$\alpha$/Fe] problem in early-type galaxies also requires AGN feedback, which plays an essential role in quenching star formation in massive galaxies where supernova (SN) feedback is inefficient \citep[hereafter \citetalias{pt15a}]{pt15a}.

For the quenching mechanism, SN-driven galactic winds have been proposed \citep{larson74,arimoto87}, but are not efficient enough for massive galaxies in hydrodynamical simulations with dark matter \citep[e.g.,][]{ck05,ck07}.
There is observational evidence for galactic winds both locally and at high redshifts, from low-mass star-forming galaxies \citep[e.g., M82,][]{ohyama02} to AGN-hosting massive galaxies \citep[e.g., Centaurus A,][]{kraft09}.
Supernova-driven winds typically have velocities of a few $10^2-10^3$ km\,s$^{-1}$ and an outflow rate comparable to the star formation rate \citep[e.g.,][]{heckman00,pettini00}, 
while AGN-driven winds show much higher velocities and outflow rates.
The presence of outflows in luminous quasars and AGN has been evidenced in broad absorption lines \citep{lynds67}.
The nearest quasar Mrk 231 shows a multi-phase outflow containing ionized, neutral \citep{rupke13,teng14}, and molecular gas \citep{feruglio10,cicone12}, which may be explained by a bipolar outflow and an accretion disk.
Winds are also seen in Seyfert galaxies \citep[e.g.,][]{tombesi11,tombesi13,pounds13,pounds14}, in which velocity probably depends on the distance from the central BHs on $\sim$pc to $\sim$kpc scales \citep{tombesi13},
and in other ultraluminous infrared galaxies with velocities depending not on star formation rate but on AGN luminosity \citep{sturm11}.
At high redshifts, AGN-hosting galaxies also show evidence of outflows \citep[e.g.,][for $z=6$]{nesvadba08,harrison12,maiolino12}. 
These suggest that AGN feedback should be activated at high redshifts.
This is the case in our AGN model where super-massive BHs originate from the first stars \citepalias{pt14}.

Galactic winds are the most plausible origin of mass--metallicity relations of galaxies \citep[e.g.,][]{tinsley80}; low-mass galaxies have lower metallicities because of larger metal loss by winds (\citealt{ck07}, see also \citealt{dave11}).
These metals are distributed to the circumgalactic medium (CGM), intracluster medium (ICM), and intergalactic medium (IGM).
The detailed mechanism of this metal transport has not been well studied yet \citep[e.g.,][]{barai14}.
Direct observations of outflows from AGN { \citep[e.g.,][]{kirkpatrick11}} and of chemical abundances near AGN will provide another stringent constraint on the modelling of AGN in galaxy simulations { \citep{germain09,barai11}}.
In \citetalias{pt15a} we showed that the mass--metallicity relation is preserved when AGN feedback is included, even though the mass of metals in massive galaxies is reduced; in this Letter we will investigate how the matter and heavy elements are redistributed from galaxies to the CGM and IGM by AGN-driven winds.

{
In this Letter, we provide a new method to identify the outflowing gas in chemodynamical simulations (Section \ref{sec:obs}).
We apply this to the most massive galaxy in our cosmological simulations, and investigate two unambiguously AGN-driven outflows.
We then describe their impact on the metallicity of the host galaxy, CGM, and IGM in Section \ref{sec:metal}.
Finally, we present our conclusions in Section \ref{sec:conc}.
The resulting spacial distributions of metallicity, and the IGM metallicity, will be shown in a future work (Taylor \& Kobayashi 2015, in prep.).

}






\section{Definition of Outflows}
\label{sec:obs}

\begin{figure}
\centering
\includegraphics[width=0.47\textwidth,keepaspectratio]{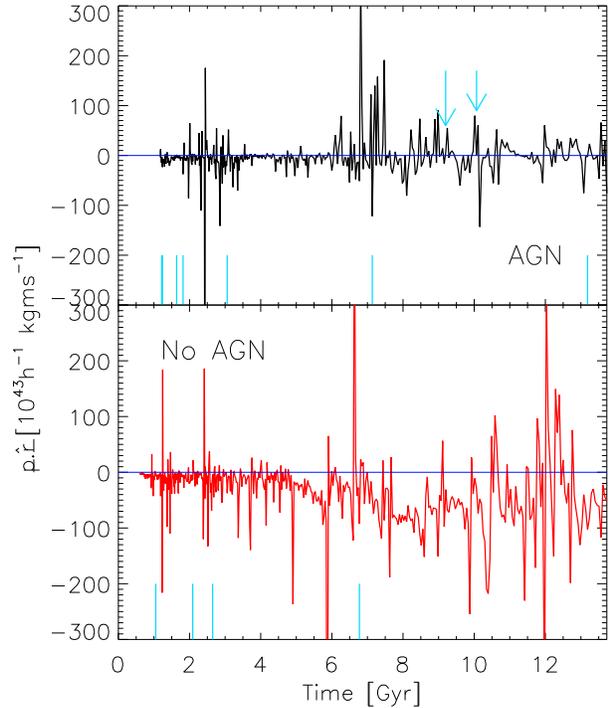}
\caption{Net radial momentum of all gas particles in a shell $20-25h^{-1}$ kpc from the centre of the largest galaxy (at $z=0$) in the simulation across cosmic time.
Positive and negative values indicate a net outflow and inflow of gas through the shell, respectively.
This value is shown for the simulation with (without) AGN feedback in the top (bottom) panel.
In both panels, blue vertical lines indicate the occurrence of major mergers.
{  The blue arrows indicate the times that the large outflows are first identified by the method of Section \ref{sec:obs}}.}
\label{fig:mom}
\end{figure}

\begin{figure}
\centering
\includegraphics[width=0.47\textwidth,keepaspectratio]{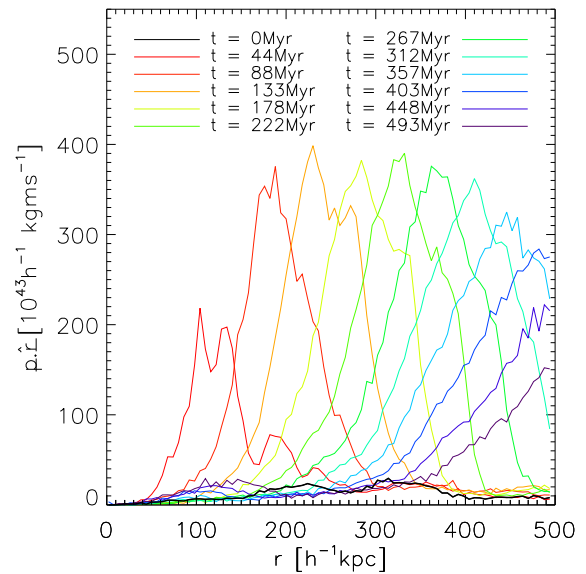}
\caption{Radial distribution of the net radial momentum of outflowing gas particles (see text for definition), within $500h^{-1}$ kpc of the most massive galaxy in the simulation with AGN.
We show this distribution at several different times {  (labelled relative to 9.201 Gyr after the Big Bang)}, allowing the progression of the outflow away from the galaxy to be seen.}
\label{fig:momrad}
\end{figure}

\begin{table*}
\centering
\caption{Total and metal masses, and abundances, of the outflows, ISM, CGM, and IGM.}
\begin{tabular}{ccccccc}
Region & $z$ & $M_{\rm g}/10^{10}M_\odot$ & $M_{\rm Z}/10^6M_\odot$ & [Fe/H] & [O/H] & [O/Fe]\\
\hline
Outflow & 0.41 & 12.3 & 268.5 & -0.78 & -0.96 & -0.09\\
Outflow & 0.31 & 8.6 & 196.6 & -0.75 & -0.94 & -0.18\\
ISM & 0.41 & 0.07 & 35.9 & 0.62 & 0.39 & -0.22 \\
ISM & 0.31 & 0.08 & 21.7 & 0.17 & 0.17 & 0.14\\
CGM (AGN) & 0.00 & 87.3 & 2302 & -0.67 & -0.78 & -0.14\\
CGM (No AGN) & 0.00 & 216 & 4292 & -0.73 & -0.96 & -0.20\\
IGM (AGN) & 0.00 & 2.08e4 & 1.14e5 & -1.77 & -1.72 & 0.00\\
IGM (No AGN) & 0.00 & 2.02e4 & 8.62e4 & -2.12 & -1.92 & 0.16\\
\end{tabular}
\label{tab:1}
\end{table*}

{ 
Our simulations use the {\sc GADGET-3} code \citep{gadget}, augmented with physical processes relevant to galaxy formation and evolution: star formation \citep{ck07}, energy feedback and chemical enrichment from Type II and Ia SNe \citep{ck04,ck09} and hypernovae \citep{ck06,ck11}; BH formation, growth through gas accretion and mergers, and energy feedback \citepalias{pt14}; heating from a uniform, evolving UV background; radiative gas cooling.
The simulation box is $25h^{-1}$ Mpc on a side, with initial conditions for $2\times240^3$ particles that produce a central cluster by the present day (see \citepalias{pt15a} for more details).
We use a gas gravitational softening length $\epsilon=1.125h^{-1}$ kpc.
}

We detect inflows and outflows by tracing gas particles for the lifetime of the simulated galaxy.
In this Letter, we focus on galaxy A \citepalias{pt15a}, which is the largest galaxy at $z=0$ in our cosmological simulations.
This galaxy sits at the centre of the cluster, and grows to $M_*=6.8\times10^{11}M_\odot$, $M_{\rm tot}=9.5\times10^{13}M_\odot$, and an effective radius of $8.1h^{-1}$ kpc by $z=0$ (see Table 1 of \citetalias{pt15a} for other quantities).
We first look at the net radial galactocentric momentum (measured in the galaxy's rest frame) of gas particles towards the outskirts of the galaxy, specifically at galactocentric radii $20h^{-1}$ kpc $<r<25h^{-1}$ kpc, for the lifetime of the galaxy.
This range was chosen to be close enough to the galaxy that gas particles passing through the shell could be associated with the galaxy, but sufficiently distant that the orbits of gas within the galaxy would not affect it.
This quantity is shown as a function of time for both simulations with and without AGN feedback in Fig. \ref{fig:mom}.
Negative values indicate a net inflow of gas; at late times, the galaxy without AGN (bottom panel) shows prolonged accretion of gas, punctuated by short-lived, SN-driven winds.
The same is not true of the galaxy when AGN feedback is included (top panel); at late times, there are a number of long-lived events in which the galaxy loses gas, most clearly between 10.5 and 12 Gyr.
In both panels, blue lines indicate the occurrence of major mergers (defined as having a mass ratio of 1:3 or greater).

Although disentangling the contributions from AGN and stellar processes is not trivial, two very massive outflows at $t\sim9.2$ Gyr ($z=0.41$) and $t\sim10.1$ Gyr ($z=0.31$) are unambiguously AGN-driven, and so we further analyse these.
In order to define the outflows, we determine which gas particles form them.
Since we are searching for AGN-driven outflows, we expect their constituent particles to be travelling radially from the galaxy centre.
Therefore, we select those particles within $500h^{-1}$ kpc of the galaxy whose radial momentum exceeds 95 per cent of its total momentum, $p$, i.e. $\boldsymbol{p\cdot\hat{r}}>0.95p$.
Choosing a slightly higher or lower threshold does not affect our results in the following sections.
Fig. \ref{fig:momrad} gives a visualization of the outflow at $t\sim9.2$ Gyr from the centre of galaxy A, showing the radial distribution, at several epochs, of the particles obtained by the above method.
Times are given relative to the time of the snapshot immediately preceding the outflow's identification, $9.201$ Gyr after the Big Bang (black line).
The progression of the outflow from the galaxy into the IGM is clear, and it can also be seen to become broader and more massive as it sweeps up other gas particles.

The total masses and metal masses of the AGN-driven outflows when they are first identified (red line of Fig.\ref{fig:momrad}) are given in Table \ref{tab:1}.
In order to decouple the outflow caused by the AGN and the (AGN-enhanced) SN wind, we ignore the particles identified from the snapshot immediately preceding it (i.e. the black line in Fig. \ref{fig:momrad}).
Those particles that remain are defined to be the AGN-driven outflow; they are hot ($T>10^7$ K) and low-density ($\sim 10^3m_{\rm H}$ m$^{-3}$).
{  On scales $\gtsim 100$ kpc, they expand isotropically with a bubble-like structure, different to the findings of \citet{costa14} who use the AREPO code; we leave a more detailed investigation of the outflow structure to a future work}.
Compared with the total gas mass $4.4\times10^{12}M_\odot$ in $R_{200}$, the outflow mass is rather small. 
However, over the next several $100$ Myr they travel further into the IGM, heating \citepalias{pt15a} and chemically enriching it (Section \ref{sec:metal}).


We also estimate the velocity of these outflows.
To each of the radial distributions in Fig. \ref{fig:momrad} we fit a Gaussian function in order to estimate its center, from which we calculate an average speed at various stages of the outflow.
These velocities are 3900 and 3400 km\,s$^{-1}$ near the galaxy for the outflows at $z=0.41$ and 0.31, respectively, falling to $\sim 1000$ km\,s$^{-1}$ at larger scales and later times.
%
%
With values for the mass and speed of the outflows, we can also estimate their kinetic energies (albeit with large uncertainties) to be $\sim 10^{60-61}$ erg. 
Note that the physical scale of our wind is much larger than for the available observations (Section \ref{sec:intro}), and our estimates of velocity and energy are larger than the observational estimates.

With our usual timestep between simulations snapshots ($\sim25-45$ Myr at these redshifts), these outflows are identified when they are already far from galactic centre ($>100$ kpc).
Therefore, we re-run the same simulation from the snapshot immediately preceding the outflow at 9.201 Gyr (i.e., from the black line to the red line in Fig. \ref{fig:momrad}), producing a larger number of snapshots with timesteps of $\sim175,000$ yr.
This allows us to trace the initial phase of the outflow, which is necessary to understand its origin.
%
The super-massive BH in this galaxy undergoes a period of accretion at 75 per cent its Eddington rate, which heats a small amount of gas close to the BH.
On such a small scale, the gas is not distributed homogeneously and isotropically around the BH, and so it is not straightforward to estimate the velocity.
If we measure the initial velocity with a kernel-weighted average of the radial velocity of the 72 gas particles nearest to the BH, we obtain $\sim8000$ km\,s$^{-1}$, which is comparable to observations of AGN-driven outflows on very small scales \citep{tombesi10,tombesi11,king13,tombesi13,pounds14}.
{  We can also estimate the mass outflow rate to be $\sim400M_\odot$\,yr$^{-1}$, which lies within the observed range $10M_\odot$\,yr$^{-1}\ltsim \dot{M}_{\rm out}\ltsim 10^3M_\odot$\,yr$^{-1}$ \citep[e.g.,][]{feruglio10,schonell14}.}
The outflowing gas soon slows and entrains more gas, eventually stalling $\sim 100$ kpc from the BH {  by $t=9.245$ Gyr, when the re-run simulation ends.}
{  The outflow seen on larger scales in the next snapshot then} suggests that there are at least 2 separate accretion events, separated by at most 88 Myr, which power the outflow seen at $z=0.41$ on the largest scales.
The AGN energy produced by this accretion can account for the kinetic energy of the large-scale outflow. 

\section{Metal Content of Outflows}
\label{sec:metal}

\begin{figure}
\centering
\includegraphics[width=0.47\textwidth,keepaspectratio]{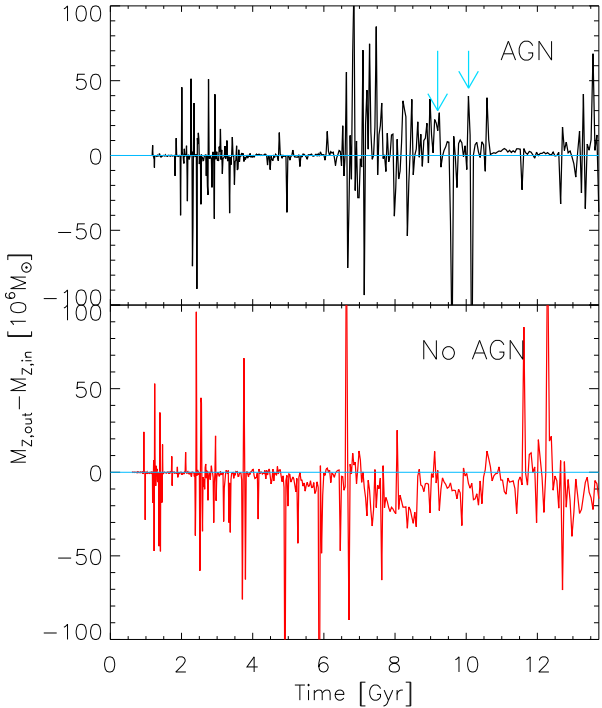}
\caption{Net flux of metal mass of all gas particles in a shell $20-25h^{-1}$ kpc from the centre of the largest galaxy at $z=0$ in the simulation, across cosmic time.
Positive and negative values indicate a net outflow and inflow of metals through the shell, respectively.
This value is show for the simulation with (without) AGN feedback in the top (bottom) panel.
{  The blue arrows indicate the times that the large outflows are first identified by the method of Section \ref{sec:obs}}.}
\label{fig:zflux}
\end{figure}

\begin{figure}
\centering
\includegraphics[width=0.47\textwidth,keepaspectratio]{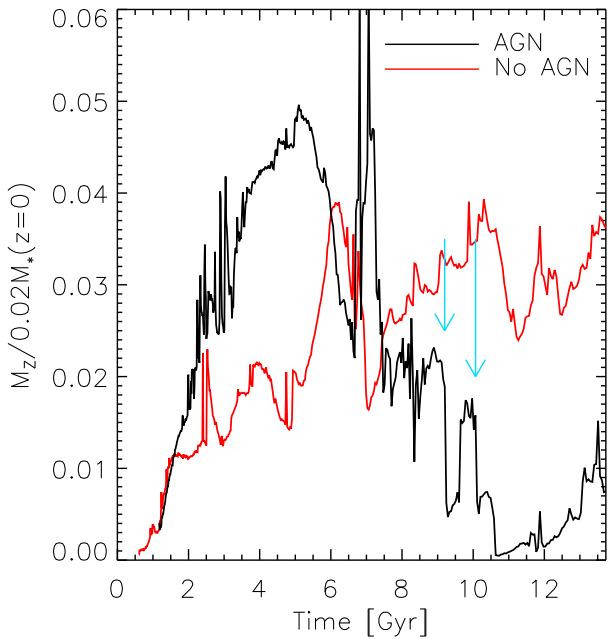}
\caption{Evolution over cosmic time of the ratio of the metal mass to the total produced metals ($=yM_*$, with the net stellar yield $y=0.02$) in the central comoving $10h^{-1}$ kpc of the galaxy.
The black (red) line indicates the simulation with (without) AGN feedback.
The blue arrows indicate the times that the large outflows are first identified by the method of Section \ref{sec:obs}.}
\label{fig:zgal}
\end{figure}

Our simulations also track the elemental abundances of gas particles, allowing us to see how the outflows affect the chemical enrichment of the host galaxy, the CGM, and the IGM.
We show in Fig. \ref{fig:zflux} the net flux of metal mass through a shell at $20h^{-1}$ kpc $<r<25h^{-1}$ kpc from the centre of the galaxy, analogous to the radial momentum flux of Fig. \ref{fig:mom}.
As in Section \ref{sec:obs}, there is a trend for net accretion of metals when AGN feedback is not included, but a net loss of metals with AGN.

This can be seen even more clearly in the top panel of Fig. \ref{fig:zgal}, in which we show the metal mass in the central comoving $10h^{-1}$ kpc of the galaxy, normalized by the total metal produced by stars (by $z=0$), as a function of time.
With AGN, the galaxy has $3.4\times10^{11}M_\odot$ of stars in its central $10h^{-1}$ kpc, while without AGN the stellar mass is $1.2\times10^{12}M_\odot$.
Although the star formation rate is lower at all times (Fig. 4 of \citetalias{pt15a}), initially the metal mass is larger with AGN feedback, {  since star formation and SN feedback is suppressed early in the life of the galaxy, and therefore a lower fraction of metals are ejected compared to the simulation without AGN feedback in which SNe alone are able to eject some metals.}
At later times, after the BH has grown to be super-massive, strong AGN feedback removes a much larger fraction than SNe alone.
The blue arrows in Fig. \ref{fig:zgal} denote the times that the large AGN-driven outflows occur, as determined by the method of Section \ref{sec:obs}.
It can be seen that they correspond exactly with sudden drops in the metal content of the galaxy.
In fact the values of these drops match very closely with the values given in Table \ref{tab:1}, obtained by identifying the outflowing particles at much greater radii, lending credence to the method in Section \ref{sec:obs}.

Similar drops are seen for the gas mass and the gas fraction.
At $\sim 8$ Gyr, this galaxy contains a significant amount of gas (a few $\times 10^9M_\odot$ in $10h^{-1}$ kpc).
The gas mass suddenly decreases due to the outflow at $9.2$ Gyr, but recovers quickly in only a few $\times 100$ Myr, which suggests that cold streams are not disrupted for long, even with this powerful outflow.
However, by tracing the ID numbers of gas particles, we find that most of the outflowing gas at $10.1$ Gyr does not fall back to the centre later.
At $\gtsim 10$ Gyr, the gas mass becomes lower than $10^9M_\odot$, which results in a gas fraction below one per cent, significantly lower than five per cent in the case without AGN feedback.

The metallicities of the outflowing gas are much lower than the average gas-phase metallicity of the galaxy in $10h^{-1}$ kpc (interstellar medium (ISM) in Table \ref{tab:1}).
This is because the gas {  expelled by} the BH originates outside the galaxy from cold flows that have undergone much less chemical enrichment; by tracing the particles back through the simulation to high redshift, we find that $\sim97$ per cent of the gas that constitutes the $z=0.41$ outflow originates in cold flows. 
The ISM metallicity increases slightly from 0.33 to 0.37 at the $z=0.41$ outflow, due to the removal of this low-metallicity gas.

Some of the ejected material accretes back onto the galaxy, and is present in the second large outflow $\sim 860$ Myr later.
At $t\gtsim 12$ Gyr, the metal mass of the ISM increases (Fig. \ref{fig:zgal}).
We can identify this origin by tracing the particles, and find that it is due to both in-situ star formation within the galaxy and mergers with gas-rich satellites.
As a result, the present gas-phase metal fraction is only a few per cent lower than the case without AGN feedback; this is the reason why AGN feedback does not affect very much the mass-metallicity relations \citepalias{pt15a}.

The AGN-driven outflows remove a significant fraction of the galaxy's metal mass, transporting it to the CGM and the IGM.
Table \ref{tab:1} gives the present metallicities of the CGM, measured in $r<150h^{-1}$ kpc, and of the IGM, which is calculated for all gas particles that do not belong to galaxies \citep[see][for the definition]{ck07}.
{  In this simulated galaxy, the oxygen mass of the CGM is $8.1\times10^8M_\odot$ with AGN, and $1.4\times 10^9M_\odot$ without; \citet{tumlinson11} estimate a lower limit of $\sim10^7M_\odot$ from observations.}
However, the fraction of CGM metals relative to the total produced in the Galaxy A is higher with AGN feedback (17.6 per cent and 13.2 per cent with and without AGN feedback, respectively), which results in the higher metallicity, $\log Z/Z_\odot=-0.78$.
%
%
With AGN feedback, 18.4 per cent of the total produced metals in the simulation box is in the IGM at $z=0$, compared with only 9.6 per cent when it is not included.
The IGM metallicity is also higher than the case without AGN, and $\log Z/Z_\odot=-1.78$; this is higher than the estimate by \citet{schaye03}, but there is a large variation in metallicities within the simulation box.
Note that, different from the simulation in \citet{ck07}, our simulation has a central concentration of galaxies, which may cause the relatively high metallicity of the IGM.

Finally, we find a difference in the elemental abundance ratio in the lower redshift outflow (see Table \ref{tab:1}); [O/Fe] ratio of the outflowing gas is $-0.18$, which is lower than for the ISM in the galaxy ([O/Fe] $= 0.14$).
The opposite trend is seen in the higher redshift outflow.
This may correspond to the selective mass-loss discussed in \citet{ck05}.
Even after star formation is quenched by the first outflow, Fe is produced by SNe Ia with a long delay.
The second outflow may be collecting this SNe Ia contribution.
On the other hand, SN-driven outflows should contain higher [$\alpha$/Fe] ratios as seen in \citet{martin02}.
Observational estimates of the metallicities and elemental abundance ratios of outflows are very important to put further constraints on the modelling of AGN feedback.


\section{Conclusions}
\label{sec:conc}

We have shown that AGN feedback drives massive outflows from the most massive galaxy in our cosmological simulations.
These outflows, {  together with metal-rich SN-driven winds,} remove gas and metals from the host galaxy, causing the CGM and the IGM to become hotter \citepalias{pt15a} and display greater chemical enrichment compared to the simulation without AGN.
Nonetheless, the mass--metallicity relation of galaxies is not affected very much by AGN feedback.

With AGN feedback, it is clear from Figs. \ref{fig:mom} and \ref{fig:zflux} that the net rate of inflow of both gas and metals into the galaxy is reduced across cosmic time. 
Focusing on the two most massive outflows detected by the net flows, we visualize the motion of outflowing gas (Fig. \ref{fig:momrad}) and quantify the AGN-driven mass and metal ejection (Table 1).
By tracing the particles, we find that most of outflowing gas originates in metal-poor cold flows, and that most of metals in outflowing gas does not fall back to the centre of the galaxy by the present.
We also find evidence for `selective' mass loss of iron relative to oxygen (Section \ref{sec:metal}).
AGN feedback is an important process not only to quench star formation at late times in massive galaxies, but also to explain chemical enrichment in circumgalactic and intergalactic medium.


\section*{Acknowledgements}
PT thanks S. Lindsay for useful discussions about identifying outflowing gas particles, and acknowledges funding from an STFC studentship.
This work has made use of the University of Hertfordshire Science and Technology Research Institute high-performance computing facility.
We thank V. Springel for providing {\sc GADGET-3}.


\bibliographystyle{mn2e}
\bibliography{./refs}



\bsp

\label{lastpage}

\end{document}